\documentclass[10pt,twocolumn,superscriptaddress,amsmath,amssymb,floats,prl,aps]{revtex4-1}
\usepackage[latin9]{inputenc}
\usepackage{amsmath}
\usepackage{graphicx}

\makeatletter

\newcommand*\LyXThinSpace{\,\hspace{0pt}}

\usepackage{comment}

\usepackage{color}
\usepackage[normalem]{ulem}

\@ifundefined{textcolor}{}{%
 \definecolor{BLACK}{gray}{0}
 \definecolor{WHITE}{gray}{1}
 \definecolor{RED}{rgb}{1,0,0}
 \definecolor{GREEN}{rgb}{0,1,0}
 \definecolor{BLUE}{rgb}{0,0,1}
 \definecolor{CYAN}{cmyk}{1,0,0,0}
 \definecolor{MAGENTA}{cmyk}{0,1,0,0}
 \definecolor{YELLOW}{cmyk}{0,0,1,0}
}

\batchmode

\@ifundefined{textcolor}{}{%
 \definecolor{BLACK}{gray}{0}
 \definecolor{WHITE}{gray}{1}
 \definecolor{RED}{rgb}{1,0,0}
 \definecolor{GREEN}{rgb}{0,1,0}
 \definecolor{BLUE}{rgb}{0,0,1}
 \definecolor{CYAN}{cmyk}{1,0,0,0}
 \definecolor{MAGENTA}{cmyk}{0,1,0,0}
 \definecolor{YELLOW}{cmyk}{0,0,1,0}
}

\usepackage{epstopdf}
\usepackage{color}
\usepackage{array}
\usepackage{dcolumn}
\usepackage{bm}

\newcolumntype{L}[1]{>{\raggedright\let\newline\\\arraybackslash\hspace{0pt}}m{#1}}
\newcolumntype{C}[1]{>{\centering\let\newline\\\arraybackslash\hspace{0pt}}m{#1}}
\newcolumntype{R}[1]{>{\raggedleft\let\newline\\\arraybackslash\hspace{0pt}}m{#1}}

\newcommand{\mbf}[1]{\mathbf{#1}}

\usepackage{tikz}
\usetikzlibrary{calc,intersections}
\usetikzlibrary{shadings}
\usepgflibrary{shadings}

\makeatother

\begin{document}

\title{Emergent magnetic degeneracy in iron pnictides due to the interplay
between spin-orbit coupling and quantum fluctuations}

\author{Morten H. Christensen}

\affiliation{School of Physics and Astronomy, University of Minnesota, Minneapolis,
MN 55455, USA}

\author{Peter P. Orth}

\affiliation{Department of Physics and Astronomy, Iowa State University, Ames,
Iowa 50011, USA}

\affiliation{Ames Laboratory, U.S. DOE, Iowa State University, Ames, Iowa 50011,
USA}

\author{Brian M. Andersen}

\affiliation{Niels Bohr Institute, University of Copenhagen, Juliane Maries Vej
30, DK-2100, Denmark}

\author{Rafael M. Fernandes}

\affiliation{School of Physics and Astronomy, University of Minnesota, Minneapolis,
MN 55455, USA}

\begin{abstract}
Recent experiments in iron pnictide superconductors reveal that, as
the putative magnetic quantum critical point is approached, different
types of magnetic order coexist over a narrow region of the phase
diagram. Although these magnetic configurations share the same wave-vectors,
they break distinct symmetries of the lattice. Importantly, the highest
superconducting transition temperature takes place close to this proliferation
of near-degenerate magnetic states. In this paper, we employ a renormalization
group calculation to show that such a behavior naturally arises due
to the effects of spin-orbit coupling on the quantum magnetic fluctuations.
Formally, the enhanced magnetic degeneracy near the quantum critical
point is manifested as a stable Gaussian fixed point with a large
basin of attraction. Implications of our findings to the superconductivity
of the iron pnictides are also discussed.
\end{abstract}

\maketitle
\emph{Introduction.\textendash{}} Magnetism in the iron pnictide superconductors
remains an intensely studied subject, not least due to its impact
on unconventional superconductivity~\cite{Mazin11,scalapino12,Chubukov12},
but also as a playground for exploring unusual types of magnetic orders~\cite{lorenzana08,eremin,fernandes16}.
While early experiments reported the prevalence of a stripe spin-density
wave (SSDW) in the phase diagrams of these systems~\cite{dagotto12,dai15},
a series of recent experiments in several different compounds found
a richer behavior~\cite{Kim10,hassinger,avci14a,khalyavin14,bohmer15a,wasser15,
malletta,mallettb,meingast16,meingast16_b,allred16a,taddei16,
hassinger16,taddei17,Frandsen17,meier17}.
As optimal doping is approached, the SSDW transition temperature is suppressed to zero, signaling a putative magnetic quantum critical point (QCP). In this region, other types of magnetic orders proliferate.
Although these are characterized by the same wave-vectors
as the SSDW phase, $\mbf{Q}_{1,2}=(\pi,0),(0,\pi)$, they do not break
the tetragonal symmetry of the lattice \textendash{} hence being dubbed
$C_{4}$ magnetic phases~\cite{avci14a}. The proximity of the superconducting
dome to this peculiar regime of intertwined magnetic phases with comparable
transition temperatures raises important questions about the interplay
between magnetism, quantum fluctuations, and superconductivity. Phenomenologically,
these $C_{4}$ phases can be understood as double-\textbf{Q }configurations
corresponding to a collinear or coplanar equal-weight superposition
of $\left(\pi,0\right)$ and $\left(0,\pi\right)$ orders \textendash{}
in contrast to the single-\textbf{Q} SSDW phase, which breaks tetragonal
symmetry and is thus called the $C_{2}$ phase~\cite{fernandes16}. 

\begin{figure}[h]
\centering \includegraphics[width=1\columnwidth]{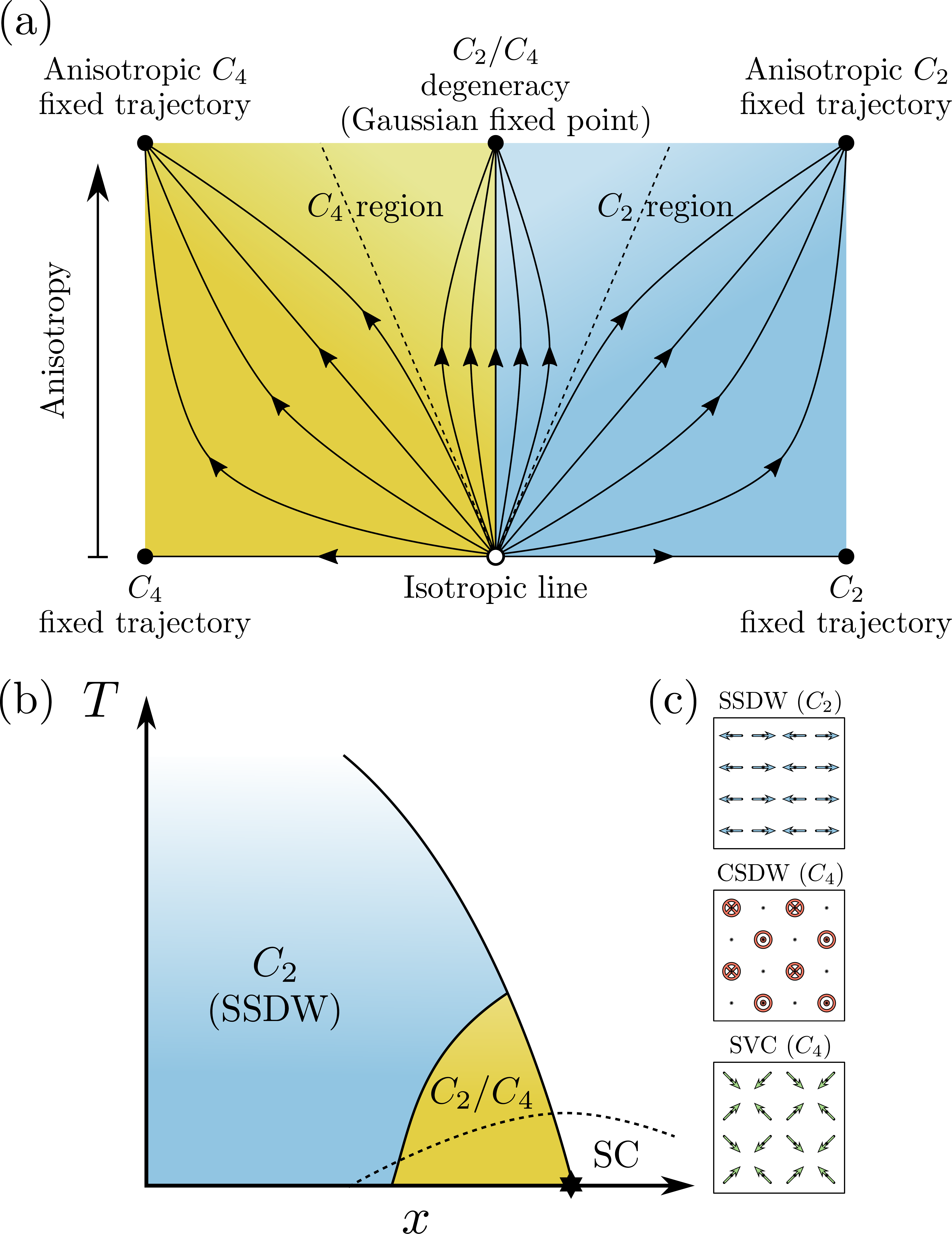}
\caption{\label{fig:schematic_diagram}(Color online) (a) Schematics of our RG results. The arrows show how fluctuations affect the coefficients of the free energy, moving them away from their mean-field values derived from a microscopic bandstructure calculation. In the spin-isotropic case, fluctuations bring the system deep into either the $C_{2}$ or a $C_{4}$ phase, removing any fine-tuned degeneracy from the system that may exist at a mean-field level. When spin anisotropies are included, a new fixed point emerges in which the $C_{2}$ and $C_{4}$ phases are degenerate. Systems whose mean-field parameters lie within the fan of dashed lines are pushed to this fixed point by fluctuations. (b) Schematic phase diagram based on our renormalization-group (RG) calculations. Here, $T$ is temperature and $x$ is an external tuning parameter, such as doping or pressure. As the putative quantum critical point (QCP, denoted by a star) is approached, the interplay between spin anisotropy (driven by the spin-orbit coupling) and quantum fluctuations leads to a near-degeneracy between the $C_{2}$ and $C_{4}$ magnetic phases. While the $C_{2}$ phase is always the stripe-spin density-wave (SSDW) state, the $C_{4}$ phase can be either the spin-vortex crystal (SVC) or the charge-spin density-wave (CSDW) state depending on whether the spin anisotropies force the moments in the plane or out of the plane, respectively, see (c).}
\end{figure}

Several microscopic mechanisms have been proposed to explain their
origin~\cite{lorenzana08,eremin,giovannetti,brydon,wang14,kang15,wang15,
gastiasoro15,christensen15,hoyer16,Agterberg17,Brink17,christensen17}.
However, they generally suffer from two drawbacks. (i) The determination
of the ground state follows from a mean-field analysis, which is unlikely
to be valid near the putative QCP due to fluctuation effects. (ii) The system is assumed
isotropic in spin space. The latter is in contradiction with the sizeable
spin-orbit coupling (SOC) observed in these systems~\cite{borisenko16},
whose $100$ K energy scale is comparable to the typical magnetic
transition temperature. As a result, the spin anisotropies induced
by SOC \cite{christensen15}, which are experimentally observed by
neutron scattering and NMR~\cite{lipscombe10,qureshi12,wang13,song13,
steffens13,li11,hirano12,curro17}, cannot be neglected near the
magnetic transition. More broadly, it is difficult to attribute the
observed near-degeneracy between the $C_{2}$ and $C_{4}$ phases
only to material-specific properties, since this behavior is often seen close to the putative
quantum critical point (see schematic Fig. \ref{fig:schematic_diagram}b) and
in several different unrelated compounds, such as hole-doped BaFe$_{2}$As$_{2}$~\cite{avci14a,bohmer15a,allred16a,meingast16_b},
pressurized FeSe \cite{Bohmer18}, and electron-doped CaKFe$_{4}$As$_{4}$~\cite{meier17}.

In this paper, we argue that this behavior is not a result of fine-tuned
interactions, but instead is a universal property of the magnetism
of the iron pnictides, provided that SOC and fluctuations are taken
into account. Although this result is complementary to previous works
on this topic, it provides a significant departure from the interpretation
that the $C_{2}$-$C_{4}$ magnetic degeneracy arises solely from band structure
effects. Universal properties are naturally described in terms of
the renormalization-group (RG) approach, which we employ throughout
this paper. The RG flow describes how the system's mean-field parameters are renormalized by fluctuations. This allows us to assess the ground states and the character of the corresponding phase transitions. 

The main results of our RG analysis near the putative magnetic QCP
are shown schematically in Fig.~\ref{fig:schematic_diagram}(a). For
the isotropic system (horizontal line), the RG flow pushes the system
deep into either the $C_{2}$ or to the $C_{4}$ phase. Moreover,
the magnetic transition becomes first-order, indicating the absence
of quantum critical fluctuations. Thus, even if the mean-field parameters
(obtained e.g. from band structure calculations) place the system
close to the degeneracy between the $C_{2}$ and $C_{4}$ phases,
fluctuations strongly remove this degeneracy. On the other hand, upon
including the spin anisotropy promoted by the SOC (vertical line),
a new fixed point in the RG flow emerges, in which the $C_{2}$ and
$C_{4}$ magnetic phases are degenerate. Importantly, when
the mean-field parameters are within the basin of attraction of this
fixed point (indicated by the region between the dashed lines in Fig.~\ref{fig:schematic_diagram}(a)), the fluctuations will drive the system to the $C_{2}$-$C_{4}$
degenerate point. Indeed, the mean-field results of e.g. Refs.~\onlinecite{gastiasoro15,christensen17} lie within this region. Thus, our main point is that the experimentally observed proliferation of nearly-degenerate $C_{2}$ and $C_{4}$ magnetic phases in the iron pnictides is not a result of fine-tuning,
but an emergent universal property of the magnetism of these systems.
In the remainder of the paper, we derive these results and discuss
their implications for the superconducting state of the pnictides.

\emph{Renormalization group flow of the isotropic case.\textendash{}}
Magnetic order in the iron pnictides is characterized by two distinct
ordering wave-vectors $\mbf{Q}_{1}=(\pi,0)$ and $\mbf{Q}_{2}=(0,\pi)$
(using single iron Brillouin zone notation).
The relative orientations and amplitudes of the magnetic vector order
parameters $\mbf{M}_{i}$ allows three types of order~\cite{fernandes16},
as illustrated in Fig. \ref{fig:schematic_diagram}(c): (i) A single-\textbf{Q}
SSDW, which takes place when only one of the $\mbf{M}_{i}$ is non-zero.
This is the phase observed in most iron pnictide parent compounds
\cite{johnston10,wen11,greene10}. (ii) A collinear double-\textbf{Q}
order dubbed charge-spin density-wave (CSDW), corresponding to $\mbf{M}_{1}=\pm\mbf{M}_{2}$.
This phase is realized, e.g. in Na-doped SrFe$_{2}$As$_{2}$ \cite{allred16a}.
(iii) A coplanar double-\textbf{Q} order dubbed spin-vortex crystal
(SVC), characterized by $\mbf{M}_{1}\perp\mbf{M}_{2}$ and $|\mbf{M}_{1}|=|\mbf{M}_{2}|$.
This phase is realized in Ni-doped CaKFe$_{4}$As$_{4}$ \cite{meier17}.
Although the three types of order share the same magnetic wave-vectors,
they break distinct symmetries of the lattice: the SSDW phase is orthorhombic whereas the CSDW and SVC phases are tetragonal
~\cite{lorenzana08,gastiasoro15,fernandes16}.

\begin{figure*}
\centering \includegraphics[width=\textwidth]{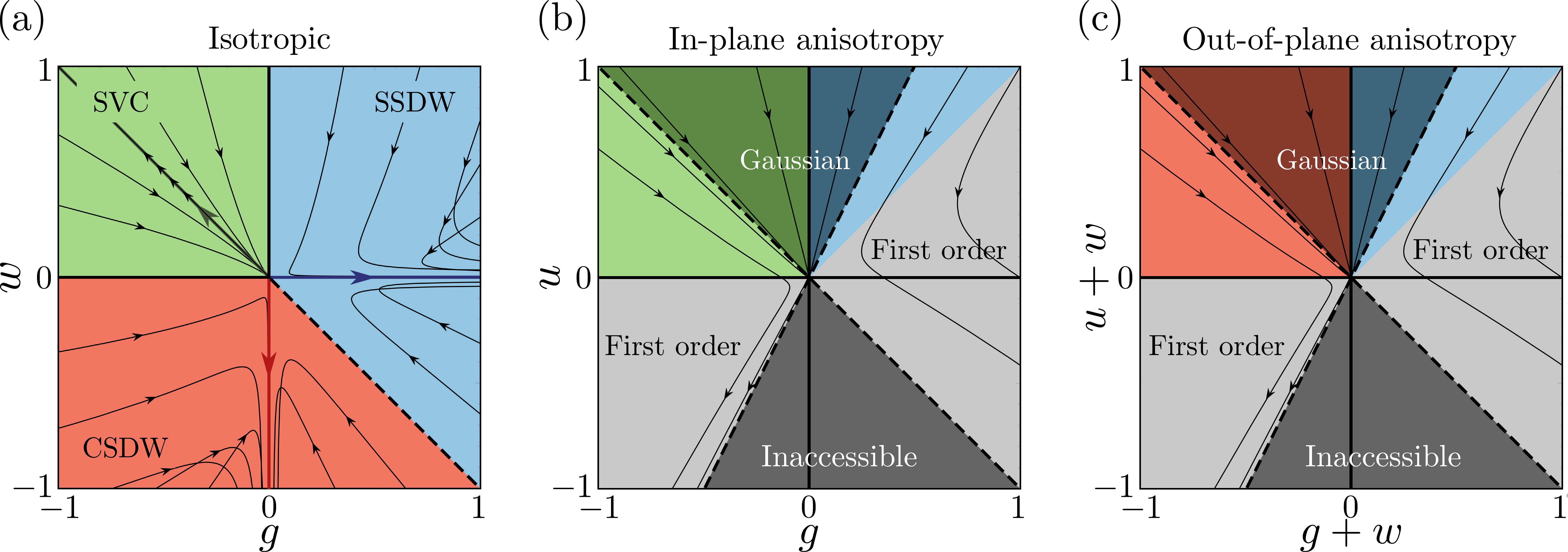}
\caption{\label{fig:all_flows} (Color online) (a) Mean-field magnetic phase diagram in the spin-isotropic case (colored background) and the RG flow lines for the zero-temperature, two-dimensional system.
There are three fixed trajectories where the quartic coefficients
diverge but their ratios remain finite. They are denoted by the thick darkly shaded lines, which lie deep inside each of the magnetic states. Because the flow lines are projections onto the $g$-$w$ plane, they appear to cross. (b)-(c) Flow diagrams for the cases of dominant (b) in-plane spin anisotropy ($\alpha_{1}<\alpha_{2},\alpha_{3}$ or $\alpha_{2}<\alpha_{1},\alpha_{3}$) and (c) out-of-plane spin anisotropy ($\alpha_{3}<\alpha_{1},\alpha_{2}$). Light gray areas
denote regions in which the free energy is unbounded, corresponding
to a first-order transition. The dark gray regions are inaccessible
to any flow. Here, darker colors denote the regions attracted to the
Gaussian fixed point, while lighter colors denote regions attracted
to the fixed trajectories. 
}
\end{figure*}

To discuss the universal properties of the magnetic phase diagram,
we introduce the magnetic action in terms of $\mathbf{M}_{1}$ and
$\mathbf{M}_{2}$ \cite{lorenzana08,eremin,giovannetti,wang15}. In
the spin-isotropic case, there are four terms allowed by tetragonal
and time-reversal symmetries:
\begin{eqnarray}
S & = &\int_{k}r_{0}(k)\left(\mbf{M}_{1}^{2}+\mbf{M}_{2}^{2}\right)+\frac{u}{2}\int_{r}\left(\mbf{M}_{1}^{2}+\mbf{M}_{2}^{2}\right)^{2}\nonumber \\
 && -\frac{g}{2}\int_{r}\left(\mbf{M}_{1}^{2}-\mbf{M}_{2}^{2}\right)^{2}+2w\int_{r}\left(\mbf{M}_{1}\cdot\mbf{M}_{2}\right)^{2}\,.\label{eq:free_energy_no_soc}
\end{eqnarray}
The coefficient of the quadratic term, $r_{0}(k)=r_{0}+\mbf{k}^{2}+\gamma|\omega_{n}|$,
is the inverse bare susceptibility, with $r_{0}\propto x-x_{c}$ denoting
the distance to the mean-field QCP $x_{c}$. Here, $\mbf{k}$ is the
momentum, $\omega_{n}=2\pi nT$ is the bosonic Matsubara frequency
with temperature $T$, and $\gamma$ is the Landau damping coefficient
arising from the decay of magnetic excitations into particle-hole
pairs. Note that $k=(i\omega_{n},\mbf{k})$ and $x=(\tau,\mathbf{r})$
with $\int_{k}\equiv T\sum_{\omega_{n}}\int\frac{\mathrm{d}^{2}k}{(2\pi)^{2}}$
and $\int_{r}\equiv\int_{0}^{1/T}\mathrm{d}\tau\int\mathrm{d}^{2}r$.
The quartic coefficient $u>0$ penalizes strong amplitude fluctuations
and ensures that the free energy is bounded. The quartic coefficient
$g$ favors either single-\textbf{Q }or double-\textbf{Q }configurations
depending on whether it is positive or negative, respectively. Similarly,
the quartic coefficient $w$ favors collinear $(w<0)$ or coplanar
($w>0$) double-\textbf{Q }configurations. The mean-field phase diagram,
obtained from straightforward minimization \cite{lorenzana08,wang14},
is shown here in Fig.~\ref{fig:all_flows}(a). 

Microscopic calculations are needed to obtain the coefficients for
specific materials. Different approaches have been proposed, from
first-principle to low-energy model calculations~\cite{lorenzana08,eremin,giovannetti,brydon,kang15,wang15,
christensen15,christensen17,gastiasoro15}.
Several of them have found regimes in which, as function of doping,
the quartic coefficients change from favoring a $C_{2}$ (i.e. single-\textbf{Q})
phase to a $C_{4}$ (i.e. double-\textbf{Q}) phase, an effect essentially
driven by changes in the band structure. Experimentally, however,
the emergence of near-degenerate $C_{2}$-$C_{4}$ phases is observed
for systems with rather different band structures, such as hole-doped
BaFe$_{2}$As$_{2}$
, electron-doped CaKFe$_{4}$As$_{4}$
, and undoped pressurized FeSe
. This suggests that the apperance
of the $C_{4}$ phase may be associated with the universal properties
of the action (\ref{eq:free_energy_no_soc}), and not only with material specific details.

To investigate this possibility, and go beyond the mean-field analysis
of previous works, we take into account the effects of fluctuations
via a renormalizaton-group (RG) calculation. In this approach, the
microscopic results discussed above provide the starting point, which
are the ``bare'' (i.e. mean-field) values of the quartic coefficients
$u_{0}$, $g_{0}$, and $w_{0}$. Upon integrating out the high-energy
magnetic fluctuations from the cutoff energy scale $\Lambda$  to the
energy $E$, these coefficients are renormalized, and become functions of the ratio $\Lambda/E$, often expressed in terms of the variable $\ell\equiv\ln\left(\frac{\Lambda}{E}\right)$. Near the putative magnetic QCP, the two-dimensional system is at its upper critical dimension, and we can use standard techniques \cite{wilson74} to derive the first-order differential RG flow equations for $u(\ell)$,
$g(\ell)$, and $w(\ell)$. The goal is to find the fixed points that
govern the critical behavior of these coefficients for a large number
of different ``initial conditions'' $u_{0}$, $g_{0}$, and $w_{0}$, corresponding to different microscopic band structures.

The RG equations of Eq.~(\ref{eq:free_energy_no_soc}) have been previously
derived~\cite{qi09,millis10,batista11,fernandes12}, but the fixed
point analysis was generally restricted to the subspace of the SSDW
phase. Our global analysis reveals that, for three-component vectors,
no stable fixed points exist. Instead, the RG flow displays three fixed trajectories, in which the quartic coefficients diverge at $\ell=\ell_{c}$, but their ratios remain fixed. They are illustrated by the colored thick
lines in Fig. \ref{fig:all_flows}(a): in all of them, $u\left(\ell\rightarrow\ell_{c}\right)\rightarrow-\infty$,
but the ratios acquire different values. The blue fixed trajectory
has $g\left(\ell_{c}\right)/u\left(\ell_{c}\right)=-1$ and $w\left(\ell_{c}\right)/g\left(\ell_{c}\right)=0$,
corresponding to a system deep inside the SSDW phase. The basin of
attraction corresponds to the blue region of the mean-field phase
diagram in Fig. \ref{fig:all_flows}(a), implying that fluctuations
do not alter the nature of the mean-field ground state. Similarly,
the two other fixed trajectories are the red line $w\left(\ell_{c}\right)/u\left(\ell_{c}\right)=1$
and $g\left(\ell_{c}\right)/w\left(\ell_{c}\right)=0$, corresponding
to a system deep inside the CSDW phase, and the green line $g\left(\ell_{c}\right)/u\left(\ell_{c}\right)=0$
and $g\left(\ell_{c}\right)/w\left(\ell_{c}\right)=-1$, corresponding
to a system deep inside the SVC phase. 

Thus, fluctuations move the system
deep into one of the ordered phases, lifting any near-degeneracy between
the $C_{2}$ and $C_{4}$ phases obtained from microscopic models
in the mean-field level. This makes it difficult to explain the proliferation
of coexisting $C_{2}$ and $C_{4}$ phases near the different optimal-doped
compounds. Furthermore, the action becomes unbounded at the fixed
trajectories, indicating a first-order quantum phase transition, and
thus no quantum critical fluctuations.

\emph{Renormalization group flow of the anisotropic case.\textendash{}}
A crucial ingredient missing in the analysis above is the spin anisotropy
that is generated by the spin-orbit coupling present in these systems~\cite{borisenko16,curro17,Li17}.
Indeed, experimentally, the magnetic moments of each configuration
are found to point to well-defined directions: in the SSDW phase,
the moments point in-plane, parallel to the wave-vector direction~\cite{dai15};
in the SVC phase, the moments also point in-plane, making $45^{\circ}$
with respect to the wave-vector directions~\cite{meier17}; in the
CSDW phase, the moments point out-of-plane~\cite{wasser15,allred16a}
[see Fig.~\ref{fig:schematic_diagram}(c)]. At the quadratic level, the
spin-orbit coupling gives rise to three different spin-anisotropic
terms~\cite{cvetkovic13,christensen15}: 
\begin{eqnarray}
 && S_{\text{SOC}}^{(2)}=\int_{k}r_{0}(k)\left(\mbf{M}_{1}^{2}+\mbf{M}_{2}^{2}\right)+\alpha_{1}\int_{k}\left(M_{x,1}^{2}+M_{y,2}^{2}\right)\nonumber \\
 && +\alpha_{2}\int_{k}\left(M_{x,2}^{2}+M_{y,1}^{2}\right)+\alpha_{3}\int_{k}\left(M_{z,1}^{2}+M_{z,2}^{2}\right)\,.\label{eq:soc_2}
\end{eqnarray}
The physical interpretation of each term is apparent: a small $\alpha_{1}$
favors in-plane moments parallel to the wave-vector directions; a
small $\alpha_{2}$ favors in-plane moments perpendicular to the wave-vector
directions; and a small $\alpha_{3}$ favors out-of-plane moments.
While the SSDW supports any of these three magnetization directions,
SVC is only compatible with the $\alpha_{1}$ and $\alpha_{2}$ terms,
and CSDW only with the $\alpha_{3}$ term. Thus, the presence of spin
anisotropies makes it impossible for the three magnetic ground states
to be nearly degenerate, but they do allow, in principle, for SSDW
to be near-degenerate with either CSDW or SVC. 

Note that the anisotropy in the quadratic terms generates anisotropies in the quartic terms. While a full solution of the RG equations is presented in Ref.~\onlinecite{christensen17b}, here we focus on limiting cases that capture the main properties of the RG flow. Due to their scaling dimension, the quadratic coefficients
$r_{0}+\alpha_{i}$ can display two
possible asymptotic behaviors as $\ell\rightarrow\ell_{c}$. Either
$r_{0}+\alpha_{i}\rightarrow\infty$, in which case the associated
spin components are quenched and do not contribute to
the action, or $r_{0}+\alpha_{i}\rightarrow-\infty$, signaling a
transition and the condensation of the spin components
related to $\alpha_{i}$. Importantly, the quadratic coefficient $\alpha_{i}$
with the smallest bare value selects which components will condense. 

Let us first consider the case of dominant in-plane anisotropy, where
initially $\alpha_{1}\left(\ell=0\right)<\alpha_{2}\left(\ell=0\right),\alpha_{3}\left(\ell=0\right)$.
The possible ground states are the SSDW phase with moments pointing
parallel to the ordering vectors and the (hedgehog)-SVC phase~\cite{meier17},
see Fig.~\ref{fig:schematic_diagram}(c). According to the discussion
above, only the components associated with $\alpha_{1}$ (namely,
$M_{x,1}$ and $M_{y,2}$) will condense, while the others can be neglected. Hence, the universal
properties of the action are the same as those of the action (\ref{eq:free_energy_no_soc})
restricted only to the $M_{x,1}$ and $M_{y,2}$ fields. As a consequence, $w$ plays no role in this case.
The RG flow of this action is shown in Fig.~\ref{fig:all_flows}(b).
Besides the two fixed trajectories equivalent to the SSDW and SVC
fixed trajectories of Fig. \ref{fig:all_flows}(b), a new
Gaussian fixed point $u_{\alpha_{1}}\left(\ell_{c}\right)=g_{\alpha_{1}}\left(\ell_{c}\right)=0$
emerges. Interestingly, we find a wide range of parameters for which
this Gaussian fixed point is attractive, indicated by the region enclosed
by the dashed lines in Fig. \ref{fig:all_flows}(b). We
note that the same phase diagram appears in the case $\alpha_{2}\left(0\right)<\alpha_{1}\left(0\right),\alpha_{3}\left(0\right)$.


In the case of dominant out-of-plane anisotropy, $\alpha_{3}\left(0\right)<\alpha_{1}\left(0\right),\alpha_{2}\left(0\right)$,
the effective action has the same form as Eq. (\ref{eq:free_energy_no_soc}),
but restricted only to the $M_{z,1}$ and $M_{z,2}$ fields. In this case, $w$ cannot be ignored although its effect can be incorporated in a shift of $u$ and $g$, as seen in the axes of Fig.~\ref{fig:all_flows}(c).
The possible ground states in this case are the SSDW and CSDW with
out-of-plane moments. As shown in Fig.~\ref{fig:all_flows}(c),
the RG flow is analogous to the case of dominant in-plane-anisotropy.
It displays two fixed trajectories corresponding to the SSDW and CSDW
states, and the Gaussian fixed point where the quartic coefficients
vanish.

The main result of our analysis is the appearance of an attractive
Gaussian fixed point in the RG flow of the anisotropic spin action.
To understand its significance, we first note that it signals a second-order
quantum phase transition (and thus quantum critical fluctuations),
in contrast to the case of fixed trajectories, which signals first-order
transitions. More importantly, at the Gaussian fixed point, the SSDW
state is degenerate with one of the $C_{4}$ phases \textendash{}
either the SVC phase for dominant in-plane anisotropy or the CSDW
phase for dominant out-of-plane anisotropy. This degeneracy is due
to the fact that, when $g=w=0$ in the action (\ref{eq:free_energy_no_soc}),
the energies of the $C_{2}$ and $C_{4}$ magnetic ground states have
the same value. The fact that the Gaussian fixed point has a wide
basin of attraction implies that, even if the bare (mean-field) values
of $g$ and $w$ are not near the phase boundary between the $C_{2}$
and the $C_{4}$ phases, quantum fluctuations will bring the system
to this special point of the phase diagram. 

\emph{Discussion.\textendash{}} Our results provide a compelling scenario
to explain the experimentally observed proliferation of $C_{4}$ phases
in close proximity to the $C_{2}$ symmetric SSDW phase as optimal
doping is approached in different iron-based compounds~\cite{Kim10,hassinger,avci14a,khalyavin14,bohmer15a,wasser15,
malletta,mallettb,meingast16,meingast16_b,allred16a,taddei16,
hassinger16,taddei17,Frandsen17,meier17}.
Instead of attributing this behavior to band structure effects, which requires fine-tuning in a wide range of compounds, our approach
reveals that the emergence of $C_{4}$ phases near the putative magnetic
QCP is a universal property of the low-energy magnetic properties
of these materials. It arises from the interplay between spin-orbit
coupling and magnetic fluctuations. We emphasize that these results
are not contradictory, but complementary to the microscopic calculations~\cite{lorenzana08,eremin,giovannetti,brydon,kang15,wang15,
christensen15,christensen17,gastiasoro15}.
In fact, our results in tandem with the mean-field results of e.g. Refs.~\onlinecite{gastiasoro15,christensen17} show that as long as the band structure effects
bring the system closer, rather than farther from the degeneracy points,
fluctuations will take over and move the system closer to the degeneracy
point. Importantly, this effect is prominent near the putative magnetic
QCP, when the system is at its upper critical dimension. This sheds
new light on why the proliferation of $C_{2}$ and $C_{4}$ phases
takes place near optimal doping, where the magnetic transition temperature
is suppressed to zero.

An important question is how this emergent $C_{2}$-$C_{4}$ near-degeneracy
impacts superconductivity. Several works have proposed an $s^{+-}$
state driven by fluctuations of the SSDW state \cite{Mazin11,scalapino12,Chubukov12}.
Usually, the existence of additional channels of magnetic fluctuations
does not guarantee an enhancement of $T_{c}$. On the contrary, in
the case of ferromagnetic \cite{Guterding17} or Néel fluctuations
\cite{Fernandes13}, they can cause pair-breaking and promote competing
superconducting states that suppress $T_{c}$ of the $s^{+-}$ state.
In our case, however, fluctuations associated with the $C_{2}$ and
$C_{4}$ phases are peaked at the same wave-vectors $\left(\pi,0\right)$
and $(0,\pi)$, and thus support the same pairing state. Therefore,
one expects that this near-degeneracy, by enhancing the phase space
of fluctuations, may cause an enhancement of $T_{c}$.
\begin{acknowledgments}
The authors are grateful to W. R. Meier, A. E. B{ö}hmer, J. Kang,
A. Kreisel, M. N. Gastiasoro, D. D. Scherer, and M. Sch{ü}tt for valuable discussions.
M.H.C. and R.M.F. were supported by the U.S. Department of Energy,
Office of Science, Basic Energy Sciences, under Award number DE-SC0012336.
B.M.A. acknowledges financial support from a Lundbeckfond fellowship
(Grant No. A9318). P.P.O. acknowledges support from Iowa State University
Startup Funds. 
\end{acknowledgments}

\end{document}